\documentclass[sigconf]{acmart}

\usepackage{booktabs} % For formal tables
\usepackage{enumitem}
\usepackage{subcaption}
\usepackage[utf8]{inputenc}

\setcopyright{rightsretained}
\acmDOI{}

\acmISBN{}

\acmConference[arXiv]{arXiv}{2017}{https://arxiv.org/} 
\acmYear{2017}
\copyrightyear{2017}

\acmPrice{15.00}

\begin{document}
\title{The Magic Barrier Revisited: Accessing Natural Limitations of Recommender Assessment}
\titlenote{Full dataset and evaluation routines available at https://jasbergk.wixsite.com/research}
%https://jasbergk.wixsite.com/research}
%\subtitlenote{}

\author{Kevin Jasberg}
\affiliation{%
  \institution{Web Science Group\\Heinrich-Heine-University Duesseldorf}
  \city{Duesseldorf} 
  \state{Germany} 
  \postcode{45225}
}
\email{kevin.jasberg@uni-duesseldorf.de}

\author{Sergej Sizov}
\affiliation{%
  \institution{Web Science Group\\Heinrich-Heine-University Duesseldorf}
  \city{Duesseldorf} 
  \state{Germany} 
  \postcode{45225}
}
\email{sizov@hhu.de}

\renewcommand{\shortauthors}{}

%\pagenumbering{arabic}
%\pagestyle{plain}
%\thispagestyle{plain}
\begin{abstract}
Recommender systems nowadays have many applications and are of great economic benefit. Hence, it is imperative for success-oriented companies to compare different of such systems and select the better one for their purposes.
To this end, various metrics of predictive accuracy are commonly used, such as the Root Mean Square Error (RMSE), or precision and recall. All these metrics more or less measure how well a recommender system can predict human behaviour.
Unfortunately, human behaviour is always associated with some degree of uncertainty, making the evaluation difficult, since it is not clear whether a deviation is system-induced or just originates from the natural variability of human decision making. At this point, some authors speculated that we may be reaching some Magic Barrier where this variability prevents us from getting much more accurate \cite{Herlocker, Hill, MagicBarrier1}.
In this article, we will extend the existing theory of the Magic Barrier \cite{MagicBarrier1} into a new probabilistic but a yet pragmatic model. In particular, we will use methods from metrology and physics to develop easy-to-handle quantities for computation to describe the Magic Barrier for different accuracy metrics and provide suggestions for common application.
This discussion is substantiated by comprehensive experiments with real users and large-scale simulations on a high-performance cluster.
\end{abstract}


% We no longer use \terms command

\keywords{Magic Barrier, Noise, Uncertainty, Distribution-Paradigm, Point-Paradigm, RMSE}
\maketitle

\section{Introduction}
Recommender systems have become quite essential for our modern information society. Applied within a variety of engines, they predict human behaviour (e.g. ratings a user might give to a specific item) and thus models a user's preferences. In doing so, those algorithms use specific machine learning techniques to learn about one's personal interests and to develop empathy for multiple as well as variable human aspects. 

Unfortunately, human beings can not be deemed as constant functions. It has recently been shown, that users provide inconsistent ratings when requested to rate same films at different times \cite{Hill}. This \textbf{Human Uncertainty}, as we understand it in this contribution, appears to be a characteristic feature of the cognitive process of decision making which influences its outcome, making it circumstantial and temporally unstable; the outcome appears to be more or less fluctuating randomly when repeating a decision making. Consequently, we may assume that observed decisions are drawn from individual distributions \cite{delia}. 

Accordingly, this complicates the evaluation of recommender systems, since it is not clear whether the difference between a given rating and the prediction is induced by the system or just a matter of Human Uncertainty. If we are able to improve the system-induced prediction quality to such an extent that only the factor of human uncertainty is left, then all visible differences within a quality metric would only exist due to this uncertainty and may vary with each repeated rating trial. This implies that rankings of different (well improved) recommender systems would shuffle with each repetition as well, i.e sound rankings do no longer exist for excellent systems but there is an equivalence class of indistinguishable optimal systems. This leads to the assumption of some Magic Barrier where natural variability may prevent us from getting much more accurate \cite{Herlocker}. 

\begin{figure}[b]
    \centering
    \begin{subfigure}{0.3\textwidth}
        \includegraphics[width=\textwidth]{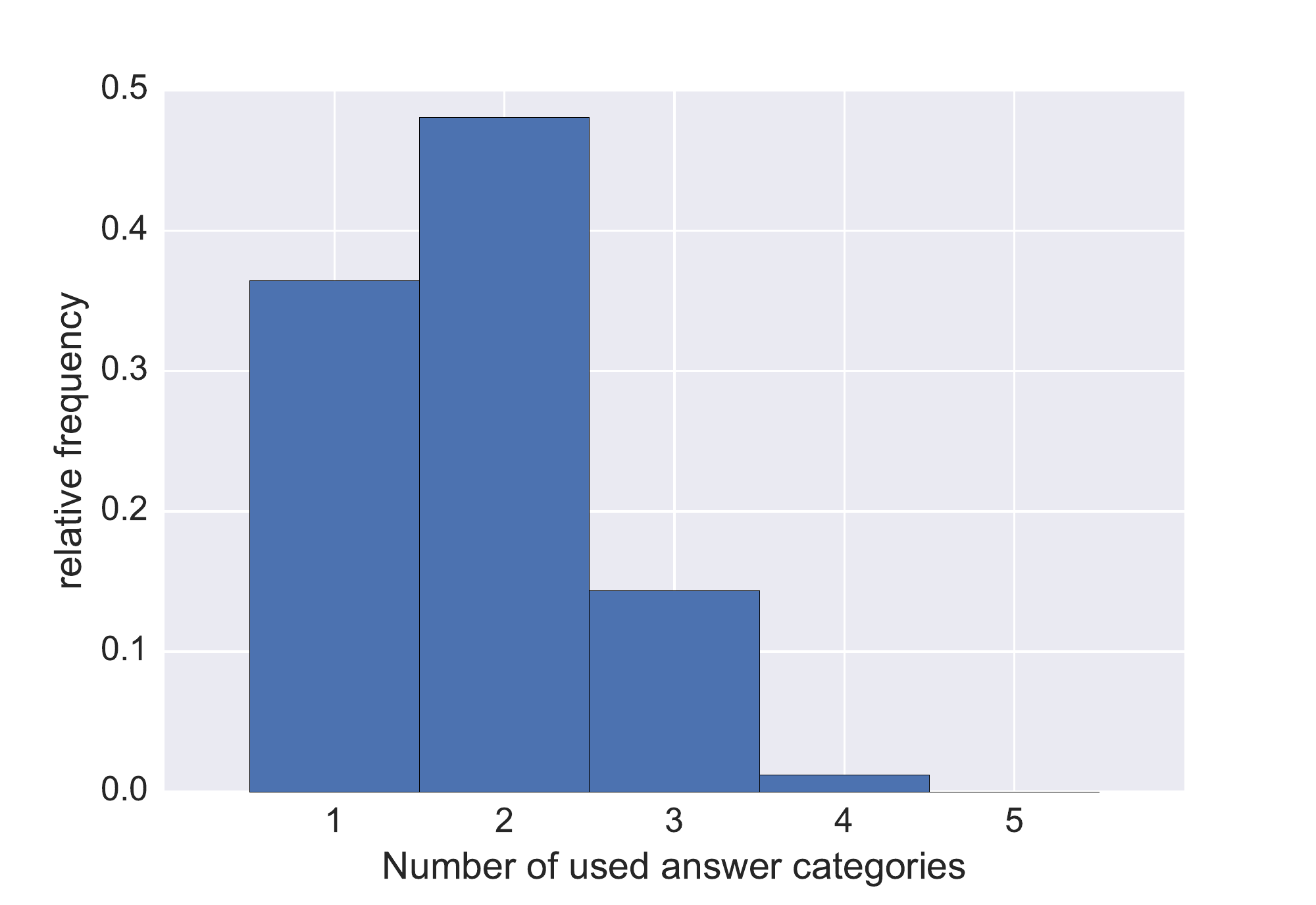}
        \caption{Frequency of used answer categories}
        \label{fig:IntroA}
    \end{subfigure}
    \begin{subfigure}{0.3\textwidth}
        \includegraphics[width=\textwidth]{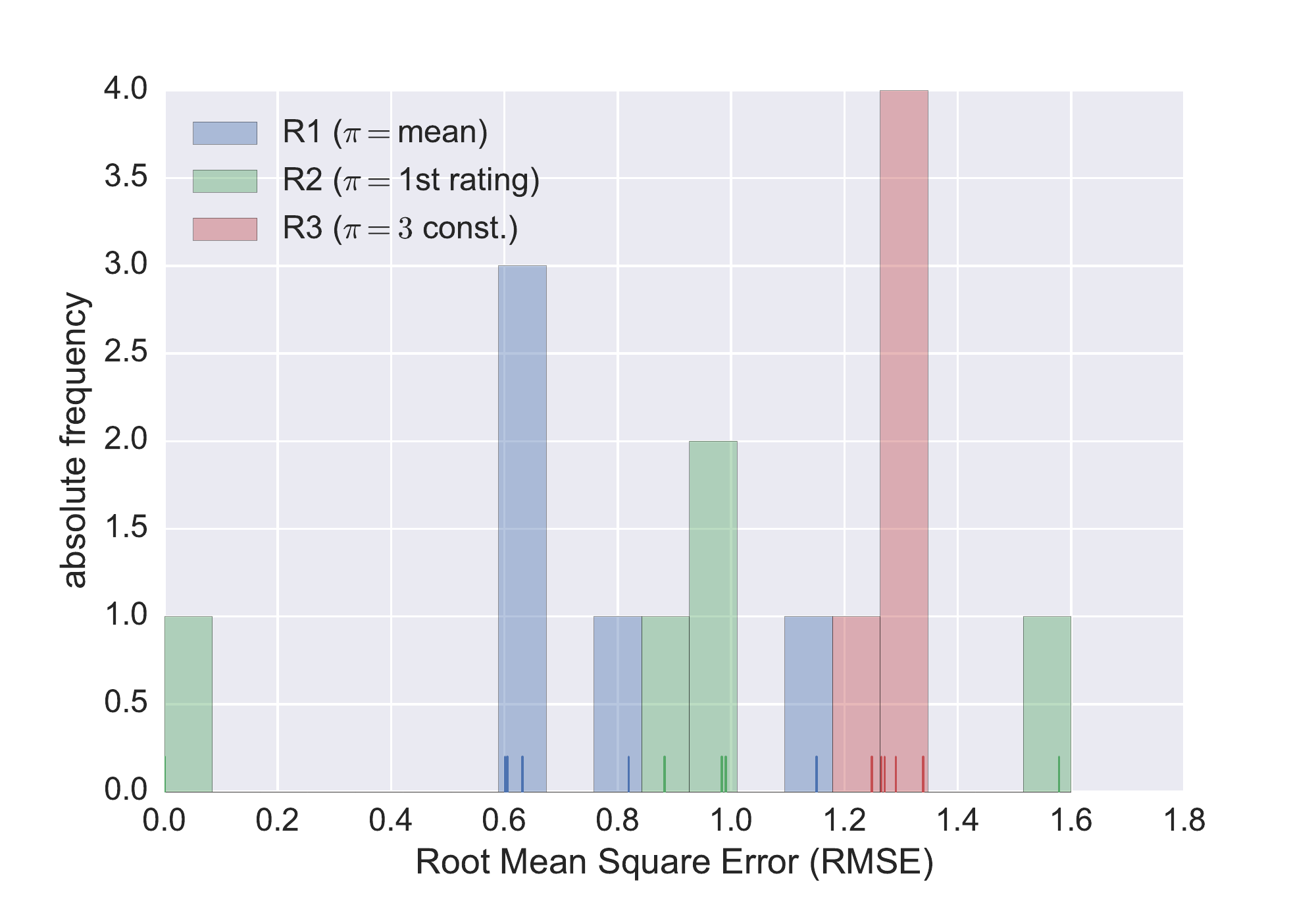}
        \caption{Distribution of RMSE outcomes}
         \label{fig:IntroB}
    \end{subfigure}
    \caption{Uncertain user ratings and impact on the RMSE}
\end{figure}

\paragraph{Motivating Example}
As a motivating example, we consider the task of rating prediction, along with the Root Mean Square Error (RMSE) as a widely used metric for prediction quality. In a systematic experiment with real users (described in more detail in forthcoming sections), individuals rated theatrical trailers multiple times. Figure \ref{fig:IntroA} shows that only 35\% of all users show constant rating behaviour, whereas about 50\% use two different answer categories and 15\% of all users make use of three or more categories. Based on these observations, we compute the RMSE for three recommender systems (designed by definition of their predictors $\pi$) for each rating trial. Figure \ref{fig:IntroB} depicts the RMSE outcomes and their frequency. It becomes apparent at once that the RMSE itself yields a particular degree of uncertainty, emerged from uncertain user feedback. When ranking these recommender systems, Figure \ref{fig:IntroB} allows for three possible results
\begin{equation}
(R1\prec R2\prec R3) \;\lor\; (R2\prec R1\prec R3) \;\lor\; (R1\prec R3\prec R2),
\end{equation}
where the relation $\prec$ denotes  ``better than''. 
The ranking problem is most obvious for recommender $R1$ as it could be both, the best or the worst recommender, although it operates for the same users rating the same items. In addition, it may be possible that further repetitions of ratings would lead to even more ranking possibilities. This naturally implies to deem those RMSE scores as single draws from distributions that are strongly overlapping. As will be revealed later, recommender $R1$ is the Magic Barrier itself. Therefore, our considerations above - the indistinguishability of excellent systems close to the Magic Barrier - hold even for straightforward investigations.

\paragraph{The Problem}
The problem of Human Uncertainty - if not explicitly considered - is that any improvement to an existing system or even the assessment of different systems might not be statistically sound. 
This, in particular, has financial implications when money is invested in the further development of a system but as a result, there is merely an overfitting instead of real improvements. Therefore, the crux is to recognise whether the prediction quality has really improved or is just some random artefact. So there is a need for a decision criterion whether a system still has room for improvements.
For the RMSE in particular, a criterion has recently been developed which allows for a dichotomous consideration (yes or no)\cite{MagicBarrier1}. But while the uncertainty of users is considered, its influence on the precise localisation of the Magic Barrier is negated. However, in our example (Fig \ref{fig:IntroB}) we have seen that the RMSE  (esp. the Magic Barrier) itself follows a distribution due to Human Uncertainty. As a consequence, systems with an RMSE near the ``old'' Magic Barrier might already be interfered by this Human Uncertainty and respectively, achieving an RMSE less than the ``old'' Magic Barrier does not always mean that this system is already interfered. So the question changes from ``Is the prediction quality interfered by Human Uncertainty?'' to ``How likely is it that the prediction quality is interfered by Human Uncertainty?'', which allows for more differentiated evaluation of recommender systems.

\paragraph{Our Objective}
In this contribution, we present a method by which the Magic Barrier can be estimated for any quality assessment metric.
For this purpose, we will embed the Magic Barrier into a complete probabilistic framework and deduce a pragmatic theory through complexity reduction. We aim to generate concrete and action-oriented quantities that can easily be embedded in existing approaches to recommender assessment. We also provide our data records for modelling Human Uncertainty and demonstrate its transferability using the example of Netflix Prize.

\section{Related Work}
\paragraph{Recommender Systems and Assessment}
The central role of recommender systems led to a lot of research and produced a variety of techniques and approaches. A good introduction and overview is given by \cite{Jannach, Handbook}. For the comparative assessment, different metrics are used to determine the prediction quality, such as the root mean squared error (RMSE), the mean absolute error (MAE), the mean average precision (MAP) along with many others \cite{Herlocker, Bobadilla, workshop12}. Although we exemplify our methodology in accordance with the RMSE, the main results of this contribution can be easily adopted for alternative assessment metrics without substantial loss of generality, insofar they require for (uncertain) human input.

%Dietmar Jannach et al.: Recommender Systems: An Introduction. Cambridge University Press, 2010
%Francesco Ricci et al.: Recommender Systems Handbook. Springer, 2011
%Bobadilla
%Herlocker
%workshop12

\paragraph{Dealing with Uncertainties}
The relevance of our contribution arises from the fact that the unavoidable human uncertainty sometimes has a vast influence on the evaluation of different prediction algorithms \cite{LikeLikeNot, noise2}.  The idea of uncertainty is not only related to predictive data mining but also to measuring sciences such as metrology. Recently, a paradigm shift was initiated on the basis of a so far incomplete theory of error \cite{Grabe, Buffler}. In consequence, measured properties are currently modelled by probability density functions and quantities calculated therefrom are then assigned a distribution by means of a convolution of their argument densities. This model is described in \cite{GUM}. A feasible framework for computing these convolutions via Monte-Carlo-Simulation is given by \cite{GUMsupp1}. We take this as a basis for our own modelling of uncertainty for addressing similar issues in the field of computer science. To derive a pragmatic and easy to handle theory, we will refer to the Gaussian Error Propagation which is commonly used in physics as well \cite{Ku,Bevington,Taylor}.

%Bevington, Philip R.; Robinson, D. Keith (2002), Data Reduction and Error Analysis for the Physical Sciences (3rd ed.), 
%Taylor, J. R., 1997: An Introduction to Error Analysis: The Study of Uncertainties in Physical Measurements

\paragraph{The Magic Barrier}
One of the first works addressing Human Uncertainty and its impact on recommender systems was presented in \cite{Hill}, where users have been proven to give inconsistent ratings on movies. The authors claim that it will never be possible to perfectly predict ratings and that there must exist an upper bound on rating prediction accuracy. Later, this upper bound was mentioned once again in \cite{Herlocker} and received the name Magic Barrier, which is still in use nowadays. A first calculation of the Magic Barrier can be found in \cite{MagicBarrier1}. Derived by risk function minimisation, the authors defined the Magic Barrier as the square root of the averaged user variances (gathered from repeated ratings). Even though this approach accounts for Human Uncertainty, its influence - namely the uncertainty of the Magic Barrier itself - remains unconsidered. In our contribution, we complete this theory and therefore allow a more differentiated analysis of recommender assessment.

\paragraph{Experimental Designs}
The complexity of human perception and cognition can be addressed by means of latent distributions \cite{delia}. This idea is widely used in cognitive science and in statistical modelling of ordinal data \cite{cub}. 
We adopt the idea of modelling user uncertainty by means of individual Gaussians following the argumentation in \cite{GaussModel} for constructing our individual response models. The methodology applied in our experiments is adopted from experimental psychology \cite{psycho} and works on repeating rating scenarios for same users-items-pairs as done before in \cite{RateAgain}.

\section{Modelling a Magic Barrier} 
In this section, we embed human uncertainty into a mathematical construct and introduce an approach for estimating a Magic Barrier for a given evaluation metric. Although the term ``Magic Barrier'' is related to the RMSE in particular, such a barrier does basically exist for any metric comparing (uncertain) user inputs with predicted scores. Therefore, we first develop a general framework which will then be illustrated for the RMSE as a prominent example.

\subsection{Changing Paradigms}
As mentioned above, various experiments \cite{RateAgain,Hill} along with our own have shown that users are scattering around their true value of preference. Consequently, we may assume that observed decisions are drawn from individual distributions, as a result of complex cognition processes, and influenced by multiple factors (e.g. mood, media literacy, etc.) \cite{delia}. Therefrom, a paradigm shift has to be carried out, which is similar to the recent change of perspectives on measurement errors in metrology \cite{Buffler}: Every measurable quantity that is somehow related to human cognition is no longer considered as a single point (point-paradigm) but rather as a whole interval of possible values (set-paradigm) that is somehow distributed (distribution-paradigm). In the context of this paper, we will, therefore, consider user ratings as random variables. On this basis, we develop statistical methodologies that are to be explored hereinafter.

\subsection{Composed Quantities}
Composed quantities, in this contribution, are quantities $Z$ that compute from a continuous function $Z=g(X_1,\ldots,X_n)$ of large amounts of uncertain arguments $X_i$ (random variables). Hence, $Z$ becomes a random variable itself. This reasoning can be understood heuristically: For each draw, there is a variety of possibilities for a single outcome $x_i$ of a random variable $X_i$. The outcomes $x_1,\ldots,x_n$ of all random variables altogether result into a single outcome for the composed quantity $Z$ by means of $z = g(x_1,\ldots,x_n)$. Accounting for all the possibilities for $x_1,\ldots,x_n$ (e.g. when repeating draws infinitely) will then result in a variety of possible outcomes $z$. Thus, the distribution of $Z$ emerges as a convolution of $n$ density functions with respect to the mapping $g$ \cite{GUM, GUMsupp1}.

\subsection{Magic Barrier Estimation}
The Magic Barrier is defined as the minimum of an evaluation metric when explicitly accounting for Human Uncertainty.
Therefore, we must first specify an optimal recommender by defining its predictors. Then we have to compute the probability density function of the evaluation metric which arises for this optimal recommender.

\paragraph{What is an optimal recommender?}
The choice of predictors depends on the evaluation metric and the underlying data model. We will demonstrate this by using an example. In the case of the Root Mean Square Error (RMSE)
\begin{equation} \label{eq:RMSE}
\text{RMSE} = \sqrt{\tfrac{1}{N}\textstyle{\sum_{\nu}} (X_\nu - \pi_\nu)^2},
\end{equation}
the comparison of a rating $X_\nu$ and a prediction $\pi_\nu\in\mathbb{R}$ is done via
$c(X_\nu)=(X_\nu - \pi_\nu)^2$, whose expectation reaches its minimum when
\begin{equation}
\tfrac{d}{d\pi} \textstyle{\sum_{i=0}^N} (x_i - \pi)^2 = 2\cdot \textstyle{\sum_{i=0}^N} (\pi-x_i) = 0
\;\Leftrightarrow\; \pi = \frac{1}{N}\textstyle{\sum_{i=0}^N} x_i \quad
\end{equation}
where $x_i$ denote the realisations of the random variable $X_\nu$.
Hence, the optimal recommender system with respect to the RMSE is defined by $\pi_\nu := \mathbb{E}[X_\nu]$ for each user-item-pair $\nu$.

This might be totally different when considering the Mean Absolute Error (MAE), whose primary comparison is based on the function $c(X_\nu)=\vert X_\nu - \pi_\nu \vert$, reaching a minimum for its expectation when $\pi_\nu$ is the median of $X_\nu$. The median corresponds to the expected value, only if a symmetrical distribution is chosen as the underlying data model. Consequently, when assuming all $X_\nu\sim\mathcal{N}(\mu_\nu,\sigma_\nu)$ to be normally distributed (symmetric density function), the optimal recommender system does not differ for the RMSE and the MAE respectively. Having $X_\nu\sim\Gamma(\alpha_\nu,\beta_\nu)$ being gamma-distributed instead, the optimal recommender may be different for both metrics, depending on the extent of asymmetry.

\paragraph{Monte-Carlo-Simulation}
Now having the definition of an optimal recommender system, we need to deduce the probability density function of the evaluation metric for this optimum. In theory, this is done by a convolution of all density functions $f_i$ of $X_i$, but what sounds simple at first, turns out to be quite laborious and inapplicable as demonstrated in \cite{Chan}. For this reason, metrologists typically apply statistical simulations.
In this paper we use \textbf{Monte-Carlo-Simulations} as described in \cite{GUMsupp1}: For each of our ratings $X_\nu$, we compute a sample $\mathcal{S}(X_\nu):= \{ x^1_\nu,\ldots, x^\tau_\nu\}$ of 
$\tau$ pseudo-random numbers (trials) that are drawn from a distribution (underlying data model). 
Then, we yield a sample for the evaluation metric $Z=g(X_1,\ldots,X_N)$ via 
\begin{equation}\label{eq:RMSE_MCM}
\mathcal{S}(Z) =\left\lbrace		z_k =  g(x^k_1,\ldots, x^k_N) \colon k=1,\ldots,\tau		\right\rbrace.
\end{equation} 
Post hoc illustration of this sample by a normed histogram with $b$ bins leads to an approximation for the density of $Z$. 

Although the statistical simulation of convolutions produces excellent results while also being easy to realise, we are facing a blatant run-time problem as soon as we are entering the realm of big data. For example, for $N = 80\,000$ ratings, the simulation already takes up to an hour of runtime\footnote{Mac mini, i5 processor, 8GB DDR3-RAM}. To compute the Magic Barrier on the Netflix test record ($N = 2.8\cdot 10^6$), we need about 35 hours.
In the following sections, we will derive a pragmatic estimate for the desired density function of the Magic Barrier for arbitrary metrics. With this, we get same results but need only a mere fraction of the simulation runtime. For example, the probability density for the Magic Barrier on the Netflix test record can be computed in less than 80 milliseconds.

\paragraph{Estimation Analytics}
Even before the technical possibilities of statistical simulations existed, metrologists had estimated the expected value and the variance of quantities $Z=g(X)$. The core these estimations is to expand $g\in C^\infty(\mathbb{R})$ into its Taylor series
\begin{equation}
g(X) = \sum_{k=0}^\infty \frac {g^{(k)}(\mu)}{k!} \, (X-\mu)^k
\end{equation}
where $g^{(k)}(\mu)$ denotes the $k^\text{th}$ derivative of $g$ evaluated at the expectation of $X$. Due to the linearity of the expectation\footnote{$\mathbb{E}[aX+b] = a\mathbb{E}[X]+b$ holds for $a,b\in\mathbb{R}$ and arbitrary random variable $X$}, we yield
\begin{eqnarray} \label{eq:TaylorExpect}
\mathbb{E}[g(X)]  
&=&  \mathbb{E} \left[ \sum_{k=0}^\infty \frac {g^{(k)}(\mu)}{k!} \, (X-\mu)^k  \right]  \nonumber
= \sum_{k=0}^\infty \frac {g^{(k)}(\mu)}{k!} \mathbb{E}\left[(X-\mu)^k\right] \\
&=&\sum_{k=0}^\infty \frac {g^{(k)}(\mu)}{k!} m_k 
\end{eqnarray}
where $m_k$ is the $k$-th central moment. For the variance and its quasi-linearity\footnote{$\mathbb{V}[aX+b] = a^2 \mathbb{V}[X]$ holds for $a,b\in\mathbb{R}$ and arbitrary random variable $X$}, we yield 
\begin{small}
\begin{eqnarray}\label{eq:TaylorVar}
\mathbb{V}[g(X)]  
&=&   \mathbb{V} \left[    \sum_{k=0}^\infty \frac {f^{(k)}(\mu)}{k!} \, (X-\mu)^k  \right]  \nonumber 
= \sum_{k=0}^\infty  \left(\frac{f^{(k)}(\mu)}{k!}\right)^2 \mathbb{V}\left[(X-\mu)^k\right] \\
&=& \sum_{k=0}^\infty \left(\frac{f^{(k)}(\mu)}{k!}\right)^2 (m_{2k}-m_k^2)
\end{eqnarray}
\end{small} 
\hspace{-1ex} where the last line has been simplified by using the common identity 
$\mathbb{V}[(X-\mu)^k] =  \mathbb{E}[(X-\mu)^{2k}] - \mathbb{E}[(X-\mu)^k]^2 = m_{2k}-{m_k}^2$. The usual approximation is to omit terms of higher orders, like
\begin{eqnarray*}
\mathbb{E}[g(X)]  &=&  g(\mu) + g'(\mu) \cdot m_1 + \ldots \approx g(\mu) \\
\mathbb{V}[g(X)]  &=& g'(\mu)^2 m_1 + g''(\mu)^2 (m_4-m_2^2)/4+  \ldots \approx  g'(\mu)^2 m_1.
\end{eqnarray*}

We have so far only considered a smooth function with just one argument in order to guarantee an easy understanding of the methodology. When considering $n$ arguments, we use a Taylor series in more dimensions and yield equivalent results which, together with the assumption of normality, form the Gaussian Error Propagation \cite{Ku,Bevington,Taylor}.

\section{Magic Barrier for the RMSE} 

\subsection{Application of Gaussian Error Propagation}

In this section we will derive closed form approximations for the RMSE and therefore define 
\begin{equation}
\mathcal{MB} = g(X_1,\ldots,X_N) := \sqrt{\tfrac{1}{N}\textstyle{\sum_{\nu}} (X_\nu - \mathbb{E}[X_\nu])^2}.
\end{equation}
Since we have to face multiple arguments, we would usually need a Taylor series in several variables, which is quite ugly for demonstration purposes. Therefore, we first condense all ratings $X_1,\ldots,X_N$ into a single random variable and then use the one-dimensional Taylor approximation. 
In doing so, we choose Gaussians as the underlying data model for our ratings. By this means, every rating $X_\nu \sim\mathcal{N}(\mu_\nu,\sigma_\nu)$ can be written as 
$X_\nu = \sigma_\nu\mathbb{I}+\mu_\nu$ where $\mathbb{I}\sim\mathcal{N}(0,1)$. 
Hence, $Y_\nu:=(X_\nu - \mathbb{E}[X_\nu])^2$ receives the expectation
\begin{eqnarray}
\mathbb{E}[Y_\nu] &=& \mathbb{E}[(\sigma_\nu\mathbb{I}+\mu_\nu -\mu_\nu)^2] 
								=     \mathbb{E}[(\sigma_\nu\mathbb{I})^2] \nonumber\\ 
								&=& \mathbb{E}[\sigma_\nu^2\mathbb{I}^2]
								=      \sigma_\nu^2 \mathbb{E}[\mathbb{I}^2]
								=      \sigma_\nu^2 \mathbb{V}[\mathbb{I}]
								=      \sigma_\nu^2 
\end{eqnarray}
as well as the variance
\begin{eqnarray}
\mathbb{V}[Y_\nu] &=&		\mathbb{V}[(\sigma_\nu\mathbb{I}+\mu_\nu -\mu_\nu)^2] 
								=     		\mathbb{V}[(\sigma_\nu\mathbb{I})^2] \nonumber\\ 
								&=& 		\mathbb{V}[\sigma_\nu^2\mathbb{I}^2]
								=      	\sigma_\nu^4 \mathbb{V}[\mathbb{I}^2]
								=     \sigma_\nu^4 \left(\mathbb{E}[\mathbb{I}^4]-\mathbb{E}[\mathbb{I}^2]^2 \right)\nonumber\\
								&=&  	\sigma_\nu^4 \left(3\mathbb{V}[\mathbb{I}]^2-\mathbb{V}[\mathbb{I}]^2 \right)
								=			2\sigma_\nu^4
\end{eqnarray}
We thus obtain a $\chi^2$-distribution for $Z:=\tfrac{1}{N}\sum_{\nu}Y_\nu$ which converges into a Gaussian for a large number $N$ of ratings by means of the central limit theorem. The parameters of this Gaussian are
\begin{eqnarray}
\mathbb{E}[Z] &=& \mathbb{E}\left[\frac{1}{N}\sum_{\nu}Y_\nu \right] 
								= \frac{1}{N}\sum_{\nu}\mathbb{E}[Y_\nu] 
								= \frac{1}{N}\sum_{\nu} \sigma_\nu^2 					\\
\mathbb{V}[Z] &=& \mathbb{V}\left[\frac{1}{N}\sum_{\nu}Y_\nu \right] 
								= \frac{1}{N^2}\sum_{\nu}\mathbb{V}[Y_\nu] 
								= \frac{2}{N^2}\sum_{\nu} \sigma_\nu^4.
\end{eqnarray}

Now we can consider the Magic Barrier to be the image of the root function of a single random variable, i.e. $\mathcal{MB} = g(X_1,\ldots,X_N) \equiv h(Z):=\sqrt{Z}$ where
$Z\sim \mathcal{N}(\frac{1}{N}\sum_{\nu} \sigma_\nu^2 \, , \, \frac{2}{N^2}\sum_{\nu} \sigma_\nu^4)$. Applying the one-dimensional Taylor approximation from equations \ref{eq:TaylorExpect} and \ref{eq:TaylorVar} leads to
\begin{eqnarray}
\mathbb{E}[\mathcal{MB}] &=& \sqrt{\mathbb{E}[Z]} - \frac{\mathbb{V}[Z] }{8\mathbb{E}[Z]^{3/2}}  - \ldots 
\approx  \sqrt{\mathbb{E}[Z]} \\
\mathbb{V}[\mathcal{MB}] &=& \frac{\mathbb{V}[Z] }{4\mathbb{E}[Z]} + \frac{\mathbb{V}[Z]^2}{32\mathbb{E}[Z]^3} +\ldots \approx  \frac{\mathbb{V}[Z]}{4\mathbb{E}[Z]}.
\end{eqnarray}
With additional assumption of normality (which is indeed a suitable model, as we will confirm soon), the approximated  distribution of the Magic Barrier for the RMSE is
\begin{equation} \label{eq:MBapp}
\mathcal{MB}\sim\mathcal{N} \left(\sqrt{\frac{1}{N} \textstyle{\sum_{\nu}} \sigma_\nu^2} \; ,\,  
\frac{1}{2N} \frac{\textstyle{\sum_{\nu}} \sigma_\nu^4}{\textstyle{\sum_{\nu}} \sigma_\nu^2}  \right)
\end{equation}
where $\mathbb{E}[\mathcal{MB}] \approx (\sum_{\nu}\sigma_\nu^2/N)^{1/2}$ exactly meets the traditional Magic Barrier as defined in \cite{MagicBarrier1} and 
$\mathbb{V}[\mathcal{MB}] \approx (\sum_{\nu}\sigma_\nu^4)/(2N\sum_{\nu} \sigma_\nu^2)$ represents the traditionally neglected uncertainty of this Magic Barrier, emerged from uncertain user ratings.

\subsection{Goodness of Approximation}
As mentioned above, the method presented here is merely an approximation, since we omit terms of higher orders. At this point, one may wonder how well this estimate actually matches the true state.
To answer this question, we first compare the simulated expectations and variances with the calculated ones in a regression analysis. Concerning the distribution model, we investigate the degree similarity using the Jensen–Shannon-Divergence.

\paragraph{Regression analysis}
We keep the following simulations as general as possible.
To this end we gradually fix a particular number $N$ of ratings from the set $\{50, 100, 150, 200, 500, 1000\}$ and sample $N$ expectations $\mu_\nu$ uniformly from the interval $[1,5]$ as well as $N$ variances $\sigma^2_\nu$ uniformly from $[\sigma^2_{min},\sigma^2_{max}]$. These intervals result from the assumption of five repeated ratings (as happened in our experiments) with the commonly used 5-star scale. Under these conditions, the positive variance yields limitations\footnote{Samples are only examples producing the minimum/maximum variance} 
\begin{eqnarray}
\sigma^2_{min} &=& \operatorname{var}(\{1,1,1,1,2\})= 0.16 \\
\sigma^2_{max} &=& \operatorname{var}(\{1,1,1,5,5\})= 3.86
\end{eqnarray}
For each pair $(\mu_\nu,\sigma^2_\nu)$ we then compute a sample $\mathcal{S}(X_\nu)$ with $\tau=10^7$ random numbers drawn from the specified Gaussian to perform the convolution via equation \ref{eq:RMSE_MCM}. For many repetitions, we receive a lot of simulated expectations/variances to be plotted against the approximated ones by means of linear regression. A perfect match between simulation and approximation would lead to the regression $y=1\cdot x+0$ with correlation coefficient $R^2=1$. The results
\begin{eqnarray}
\operatorname{Sim}(\mathbb{E}) &=& 0.999 \cdot \operatorname{Apr}(\mathbb{E}) - 0.003  \qquad (R^2=0.99)\\
\operatorname{Sim}(\mathbb{V}) &=& 0.981 \cdot \operatorname{Apr}(\mathbb{V}) + 0.000 \qquad (R^2=1.00)
\end{eqnarray}
show that this condition is almost fully achieved and hence we may consider these approximations as appropriate.

\paragraph{Jensen–Shannon-Divergence}
\begin{figure}[t]
\centering
\includegraphics[width=\linewidth]{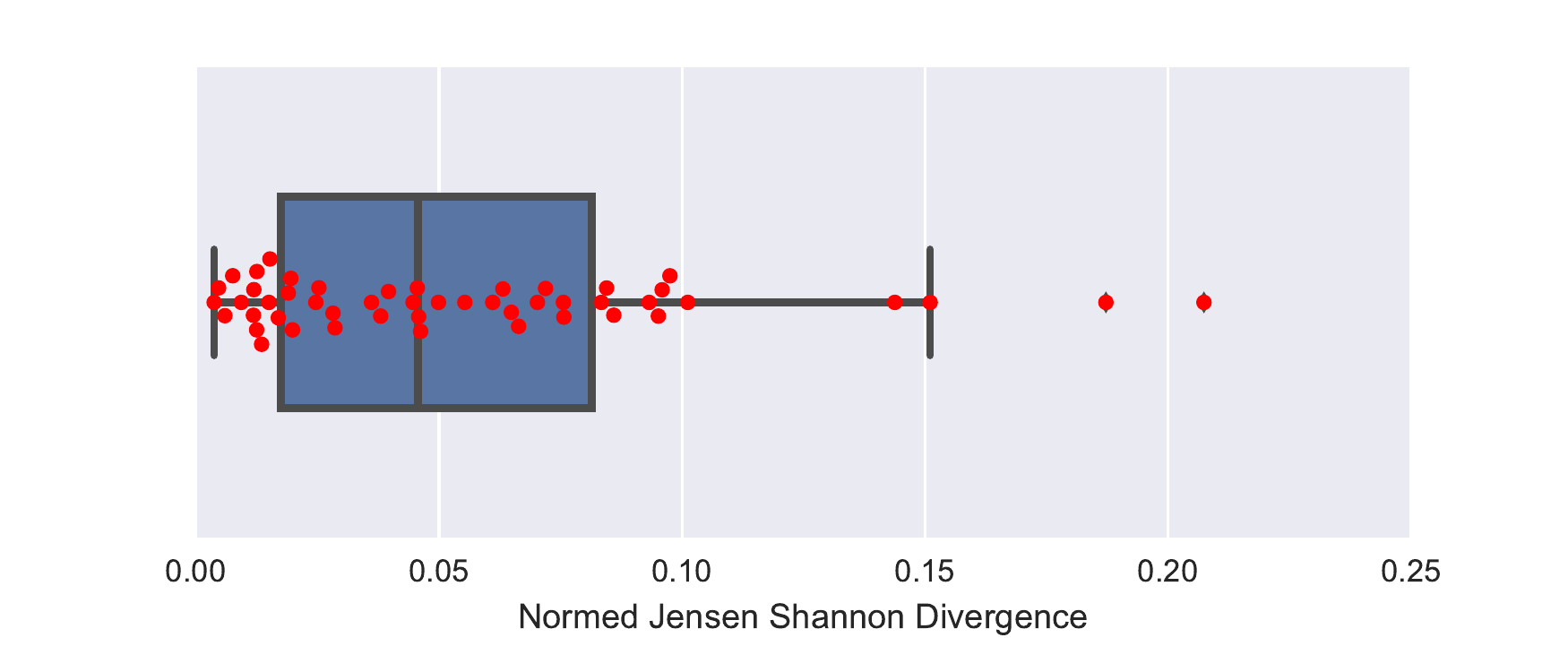}
\caption{Jensen–Shannon-Divergence for comparing the simulated distribution with a predetermined Gaussian}
\label{fig:JSD}
\end{figure}
When modelling the Magic Barrier, not only the expectation and the variance are of great importance, but rather the entire probability density. While the simulated distribution arises naturally from convolution, it is predetermined for the approximation. Therefore, it is necessary to evaluate the degree of deviation of both distributions. In doing so, we proceed as done in the regression analysis above, but instead of computing means and variances, we transform our samples into discrete probability distributions $P_{sim}$ and $P_{apr}$ and analyse the Jensen–Shannon-Divergence (JSD)
\begin{equation}
\operatorname{JSD}(P_{sim}| P_{apr}) 
= \frac{1}{2} D_\mathrm{KL}(P_{sim}| M) + \frac{1}{2} D_\mathrm{KL}(P_{apr}| M)
\end{equation}
where $D_\mathrm{KL}(P_1|P_2)=\sum _{i}P_1(i)\,\log_2 (P_1(i)/P_2(i))$ denotes the Kullback-Leibler-Divergence and  $M=\frac {1}{2}(P_{sim}+P_{apr})$. Since we use the base $2$ logarithm, the JSD yields the boundaries 
\begin{equation}
0 \leq \operatorname{JSD} \leq 2\log(2)  \quad\text{or}\quad 0 \leq \tfrac{\operatorname{JSD}}{2\log(2)} \leq 1 
\end{equation}
The outcomes for the normed JSD is shown in Figure \ref{fig:JSD}. We observe that the mid-range of all outcomes is located between $0.01$ and $0.08$ confirming high similarity of the simulated distribution and the assumed Gaussian. There are, however, some outliers which only occur for $N=50$ ratings. This can be explained by the fact that the RMSE contains the sum of squared normal distributions, which is $\chi^2$-distributed, but quickly converges to the normal distribution for $N>100$. Thus, the more ratings we have, the more adequate is a Gaussian as the assumed density. For a visual comparison of both distributions, Figure \ref{fig:ExpBarrierComp} depicts the simulated density as well as the approximation for our experiment with $N=213$.

\subsection{Understanding the Magic Barrier}
\begin{figure*}
    \centering
    \begin{subfigure}{0.24\textwidth}
        \includegraphics[width=\textwidth]{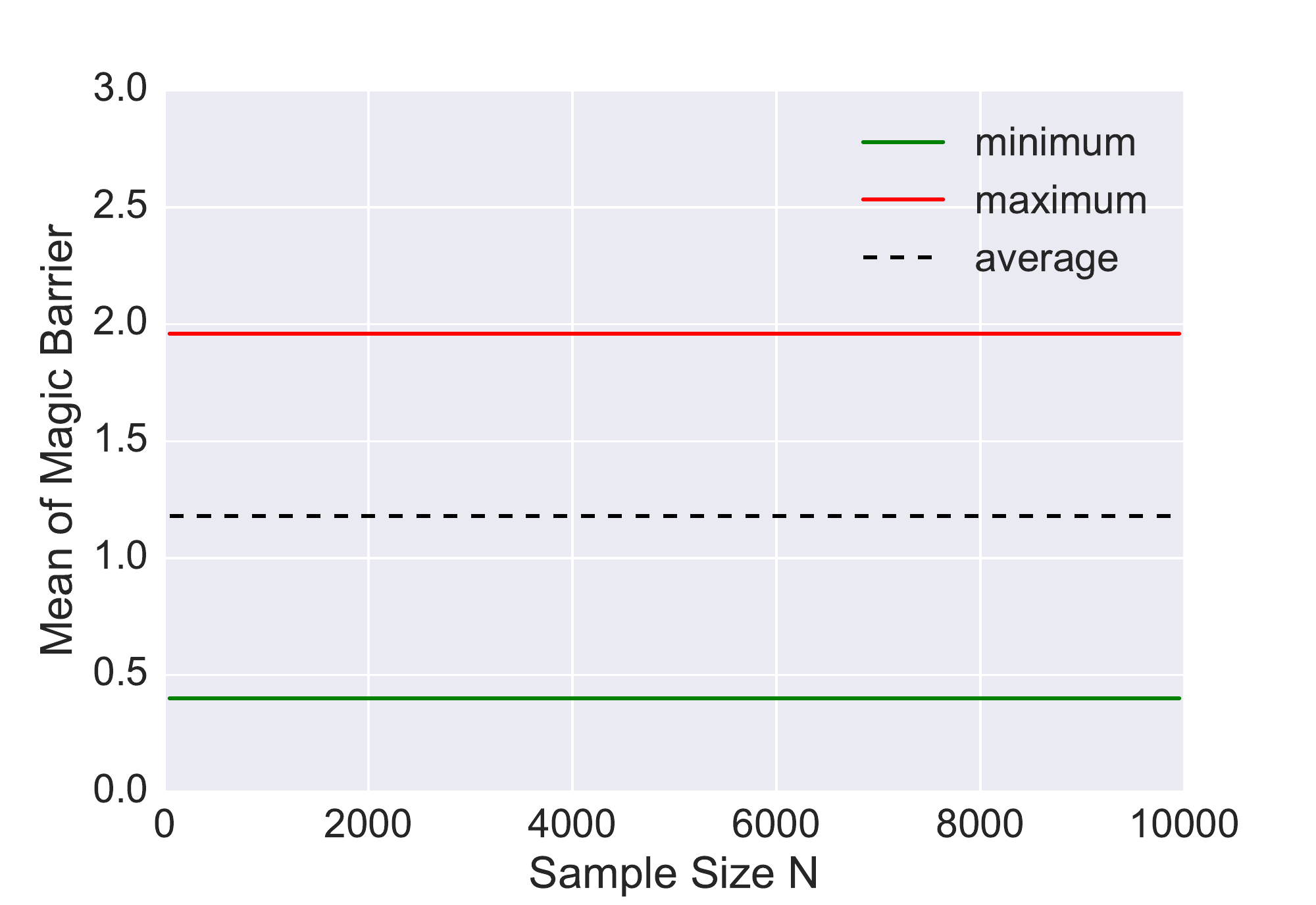}
        \caption{$\mathbb{E}[\mathcal{MB}]$ with respect to $N$}
        \label{fig:SensA}
    \end{subfigure}
    \hfill
    \begin{subfigure}{0.24\textwidth}
        \includegraphics[width=\textwidth]{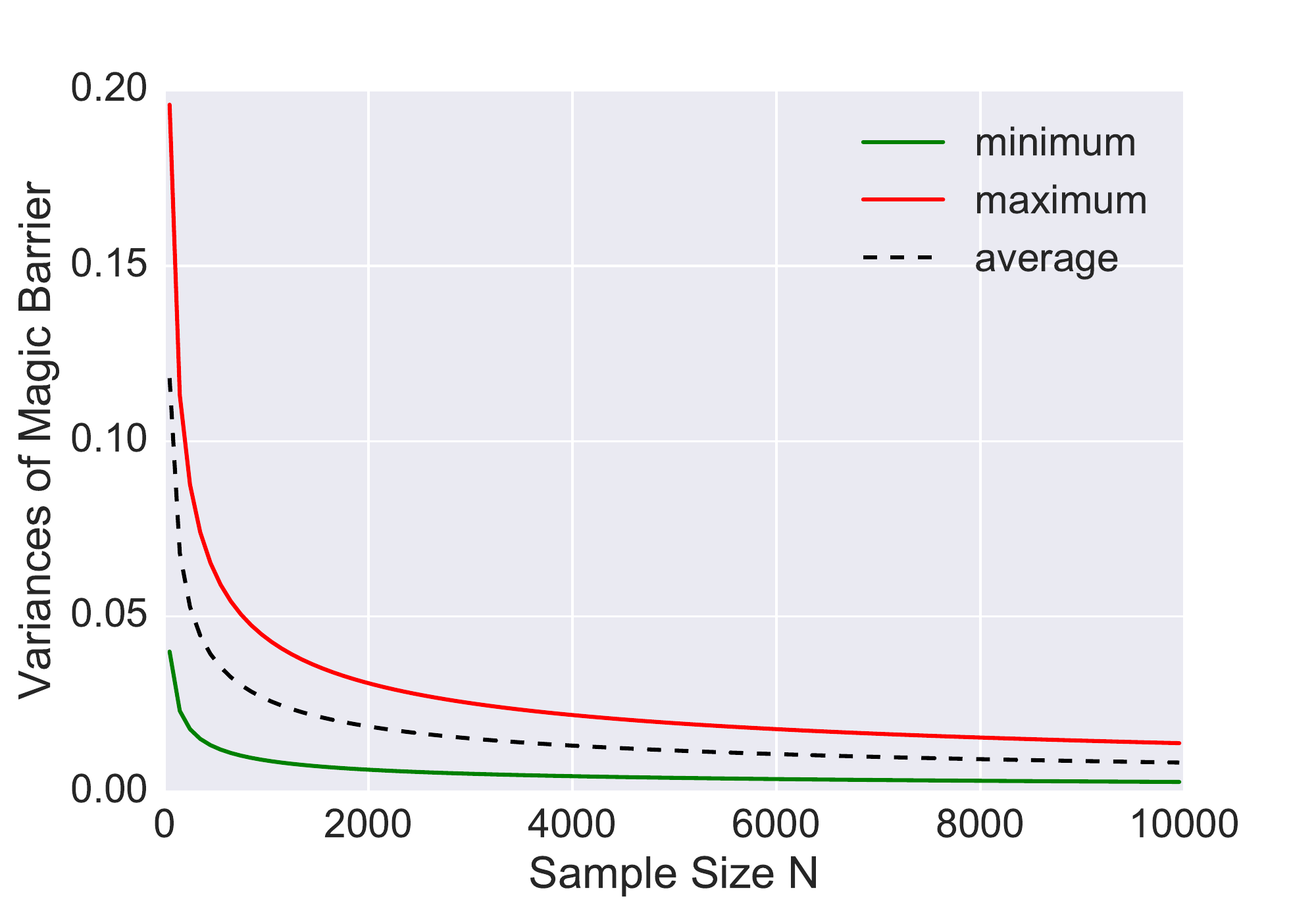}
        \caption{$\mathbb{V}[\mathcal{MB}]$ with respect to $N$}
         \label{fig:SensB}
    \end{subfigure}
     \hfill
    \begin{subfigure}{0.24\textwidth}
        \includegraphics[width=\textwidth]{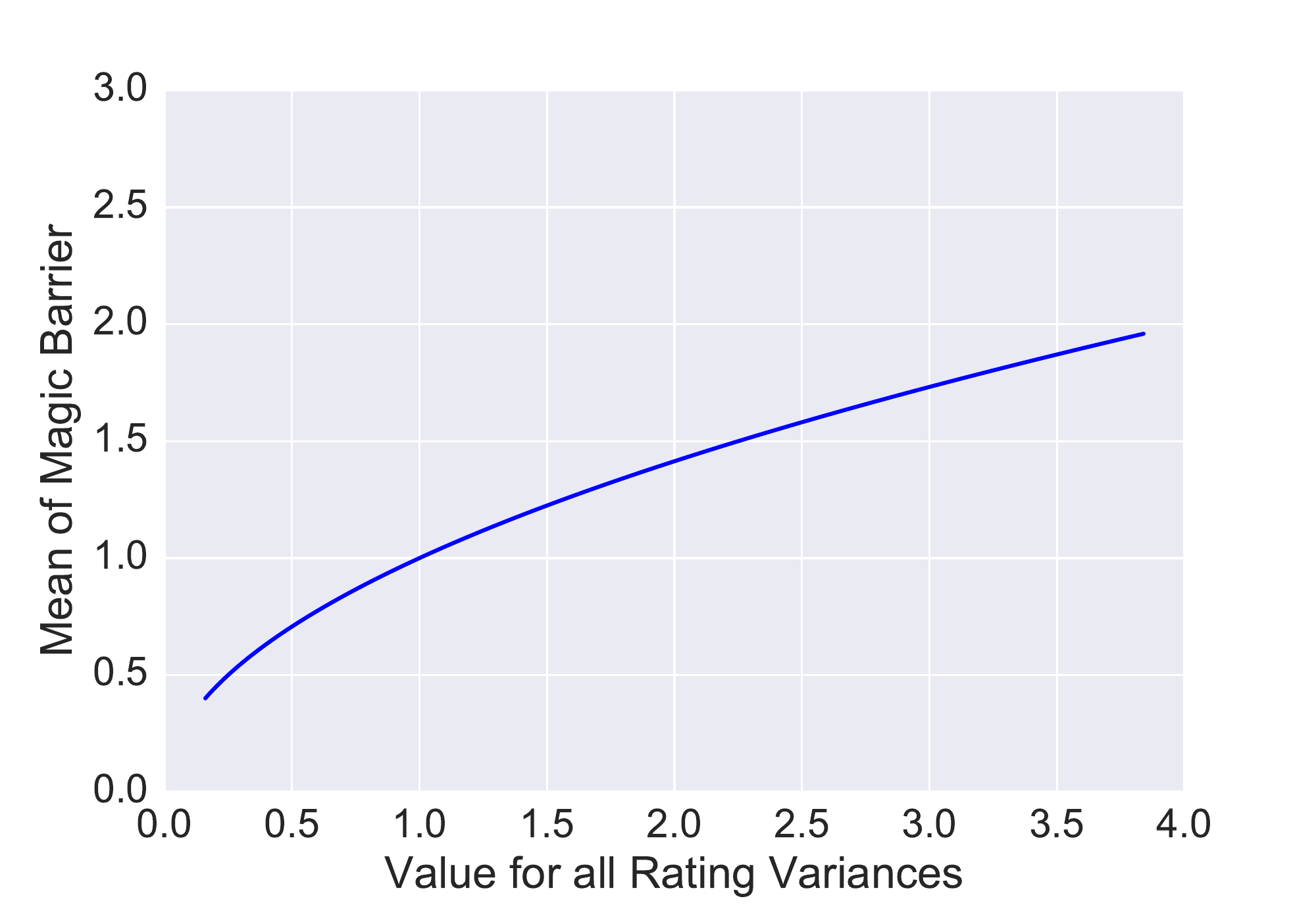}
        \caption{$\mathbb{V}[\mathcal{MB}]$ with respect to $\sigma^2$}
         \label{fig:SensC}
    \end{subfigure}
     \hfill
    \begin{subfigure}{0.24\textwidth}
        \includegraphics[width=\textwidth]{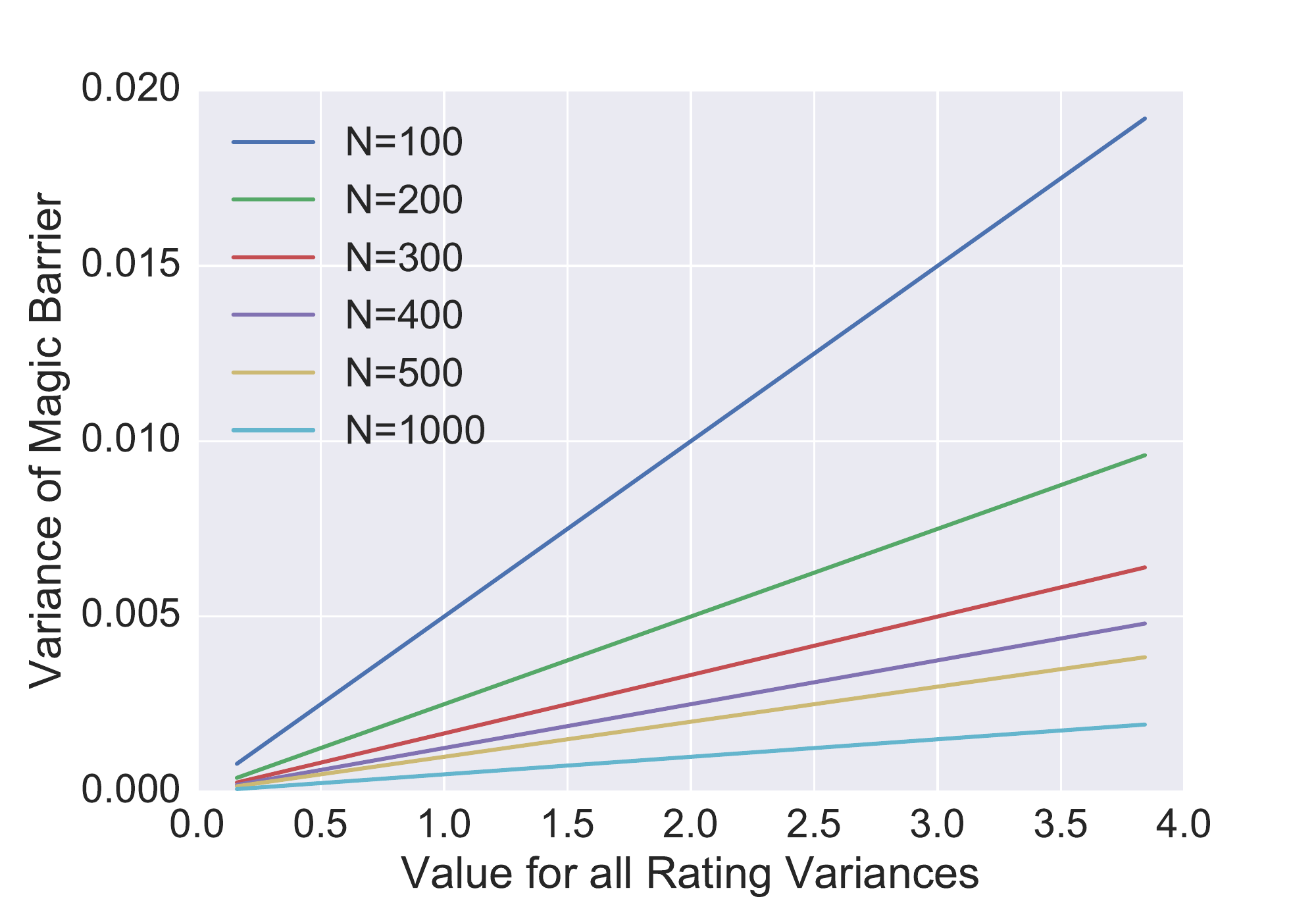}
        \caption{$\mathbb{V}[\mathcal{MB}]$ with respect to $\sigma^2$}
         \label{fig:SensD}
    \end{subfigure}
    \caption{Sensitivity analysis of the Magic Barrier varying the number of ratings and the extent of rating variances}
\end{figure*}

In this section, we will take a closer look at the properties of the Magic Barrier. For this purpose, the individual dependencies of the Magic Barrier and their effects are analysed in a sensitivity analysis.
In addition, we will generalise the dichotomous decision criterion from \cite{MagicBarrier1} and develop a pragmatic rule of thumb to ascertain whether a deeper consideration of the Magic Barrier seems worthwhile.

\paragraph{Sensitivity analysis}
A sensitivity analysis is used to determine how a quantity responds to the variation of its arguments. Therefore, we vary one argument within reasonable boundaries while fixing all the other arguments at the same time.

In Figure \ref{fig:SensA} and \ref{fig:SensB}, one can observe the Magic Barrier's reaction to an increasing number $N$ of ratings. It is seen that the expectation remains unaffected by the number of uncertain ratings. Only the extent of the uncertainty raises or lowers the mean value. On a 5-star scale together with five re-ratings, the expected value yields limitations (green and red) due to the minimum and maximum variance possible. The growth behaviour of the expectation under rating uncertainty is asymptotic. 
However, Figure \ref{fig:SensB} reveals that the Magic Barrier's variance is heavily impacted by the number of ratings, i.e. the precision of the Magic Barrier even gains when more uncertain ratings are added. The extent of rating uncertainty also leads to boundaries, but this influence gradually disappears for increasing $N$. A comparison of Figure \ref{fig:SensB} and Figure \ref{fig:SensD} reveals that the number of ratings significantly affects the first two decimal places of the variance, whereas the influence of the rating uncertainty affects only the third and fourth decimal places at most. In summary, it can be said that the extent of Human Uncertainty alone is responsible for the location of the Magic Barrier, whilst its spread can be reduced by adding ratings. However, the degree of this improvement decreases very rapidly.

What we have omitted here is the influence of the underlying data model and the applied rating scale. The rating scale limits the variance of a user and thus has a great impact of the possible location of the Magic Barrier. The underlying data model has also a great impact on the Magic Barrier but will be the discussed separately in further research.

\paragraph{Do we need a Magic Distribution?}
Now having in mind that the variance of the Magic Barrier decreases for large $N$, one may ask if we really need the Magic Barrier to be a distribution rather that a single score. The answer depends on many factors. First of all, the world of recommender assessment does not entirely consist of large-scale experiments, so that the variance can not be deemed to equal zero. In the case of large-scale experiments, the predefined accuracy of computed scores does matter quite a lot. For example, all RMSE scores were given to the fourth decimal place in Netflix Prize \cite{netflixrules}. As shown in the following sections, the standard deviation of the Magic Barrier for the Netflix data set can be assumed to be $\sigma=0.0007$, which is still seven times larger than the specified rounding accuracy of four decimal places.
In this example, we see that even for large data records the effort of considering the Magic Barrier as a distribution is quite meaningful.
\begin{figure}[b]
        \includegraphics[width=\linewidth]{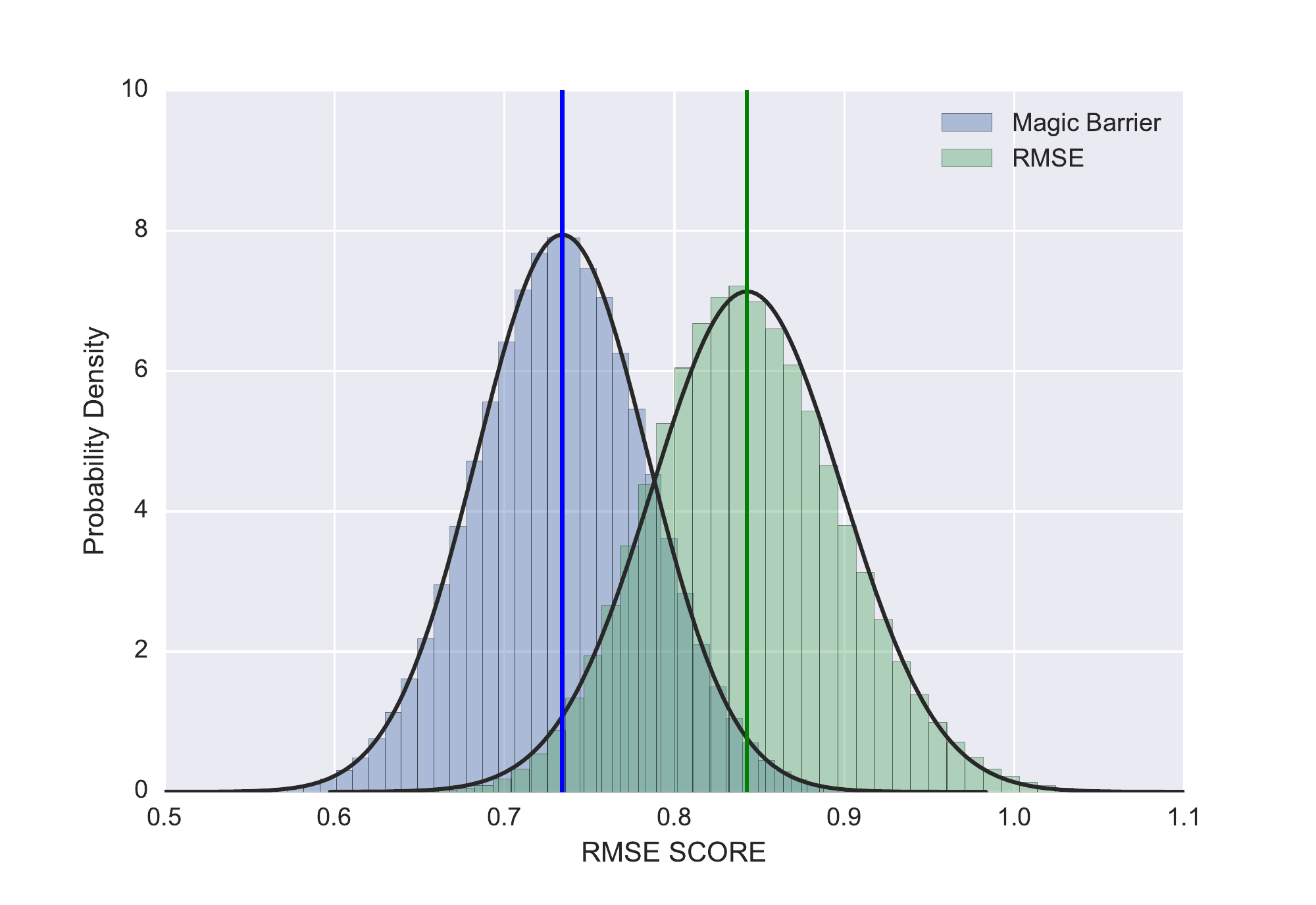}
        \caption{Interference of RMSE with the Magic Barrier}
        \label{fig:ErrExa}
\end{figure}

Furthermore, we need a non-vanishing variance for a statistically sound decision whether a system can still be improved.
Following \cite{MagicBarrier1}, any improvement of a recommender system is pointless, if the RMSE score is below the  Magic Barrier, i.e. $\mathbb{E}[\text{RMSE}]<\mathbb{E}[\mathcal{MB}]$. But since both quantities are distributed, their density functions may nevertheless overlap. Figure \ref{fig:ErrExa} illustrates the interference of the Magic Barrier with a recommender system used in our experiments. Although the decision criterion from \cite{MagicBarrier1} holds, there is a significant probability that the RMSE outcome is already affected by the Magic Barrier. This probability is given by
\begin{equation}\label{eq:BarrierInterference}
P(\mathcal{MB}>\text{RMSE}) 
= \int_{-\infty}^\infty f_{\text{RMSE}}(x)\cdot\big( 1-F_{\mathcal{MB}}(x) \big) \,\mathrm{d}x
\end{equation}
where $F_{\mathcal{MB}}(x)$ denotes the cumulative distribution function. In our example from Figure \ref{fig:ErrExa}, this probability is around $0.33$, i.e. the RMSE is interfering with  the Magic Barrier in one of three outcomes. For this reason, an analysis of possible improvements can not be answered by a dichotomous decision criterion (yes or no), but has to be answered by means of probabilities (How likely is it that my system can still be improved and what risk am I willing to accept?).

\paragraph{When is a differentiated consideration needed?}
However, such a differentiated approach is not always worth it.
Therefore, it would be useful to have a rule of thumb to find out whether a differentiated consideration is fruitful or not. For example, a possible criterion might be the intersection of the 99\%-confidence intervals of the RMSE and the Magic Barrier. Due to normality, further analysis should be taken into consideration, when
\begin{equation} \label{eq:criterion}
\mathbb{E}[\mathcal{MB}] + 3\sqrt{\mathbb{V}[\mathcal{MB}] }> \mathbb{E}[\text{RMSE}] - 3\sqrt{\mathbb{V}[\text{RMSE}]}.
\end{equation}
By assuming $\mathbb{V}[\mathcal{MB}] \approx \mathbb{V}[\text{RMSE}]$, which usually holds when both quantities are computed on the same data record, this criterion can be simplified to 
$\mathbb{E}[\text{RMSE}] - \mathbb{E}[\mathcal{MB}] < 6\mathbb{V}[\mathcal{MB}]^{1/2}$.

\section{Experiments}

In this section, we examine our theoretical considerations in reality.
To this end, we conducted a controlled experiment with real users and measured their uncertainty. We are thus able to support the chosen data model and verify our approximation on a real data set. On this basis, possible applications can be illustrated (e.g. transferring our variances to other situations where no Human Uncertainty was explicitly measured).

\subsection{The Experiment}
Our experiment is set up with Unipark's\footnote{http://www.unipark.com/de/} survey engine while our participants were committed from the crowdsourcing platform Clickworker\footnote{https://www.clickworker.de/}.
To derive a user's rating distributions, we use the method of re-rating, which was successfully used in \cite{RateAgain, Hill} before. For this purpose, participants watched theatrical trailers of popular movies and television shows and provided ratings in five repetition trials\footnote{A full description can be found on https://jasbergk.wixsite.com/research}. User ratings have been recorded for five out of ten fixed trailers so that remaining trailers act as distractors triggering the misinformation effect, i.e. memory is becoming less accurate due to interference from post-event information. 

We received a rating tensor $R_{u,i,t}$ with $\dim(R)=(67,5,5)$, having $N=1\,675$ ratings in total, where the coordinates $(u,i,t)$ encode the rating that has been given to item $i$ by user $u$ in the $t$-th trial. From this record we derive a unique rating distribution for each user-item-pair by considering tensor-slices in trial-dimension 
$R_{u,i}:= \{ R_{u,i,t} \vert t=1,\ldots ,5 \}$ for which we compute Maximum-Likelihood-Parameters given a predetermined data model (e.g. Gaussians, CUB-Models, etc.). Altogether $67$ people from Germany, Austria and Switzerland participated in this experiment. This group can be parted into $57\%$ females and $43\%$ males whose ages range from $20$ to $60$ years while over $60\%$ of our participants where aged between $20$ and $40$. This group also includes a good average of lower, medium and higher educational levels. The rating frequency habits range from ``rarely'' to ``often'' in uniform distribution.

\subsection{Data Model and Uncertainties}

\paragraph{Proving the data model}
In this contribution we opt for Gaussians since they are strongly associated to human characteristics \cite{PerceptionCognition} and have also been proven to be appropriate user models in \cite{GaussModel}. Additionally, Gaussians exhibit maximum entropy along all distributions with finite mean/variance and support on $\mathbb{R}$.  

For each recorded item, all tensor slices having a non-vanishing variance are checked for normality by means of a one-sample KS-test \cite{KS} with confidence level $\alpha=0.05$. The null hypothesis was never rejected, allowing to keep the Gaussian distribution as a possible model.

\begin{figure}[b]
        \includegraphics[width=.9\linewidth]{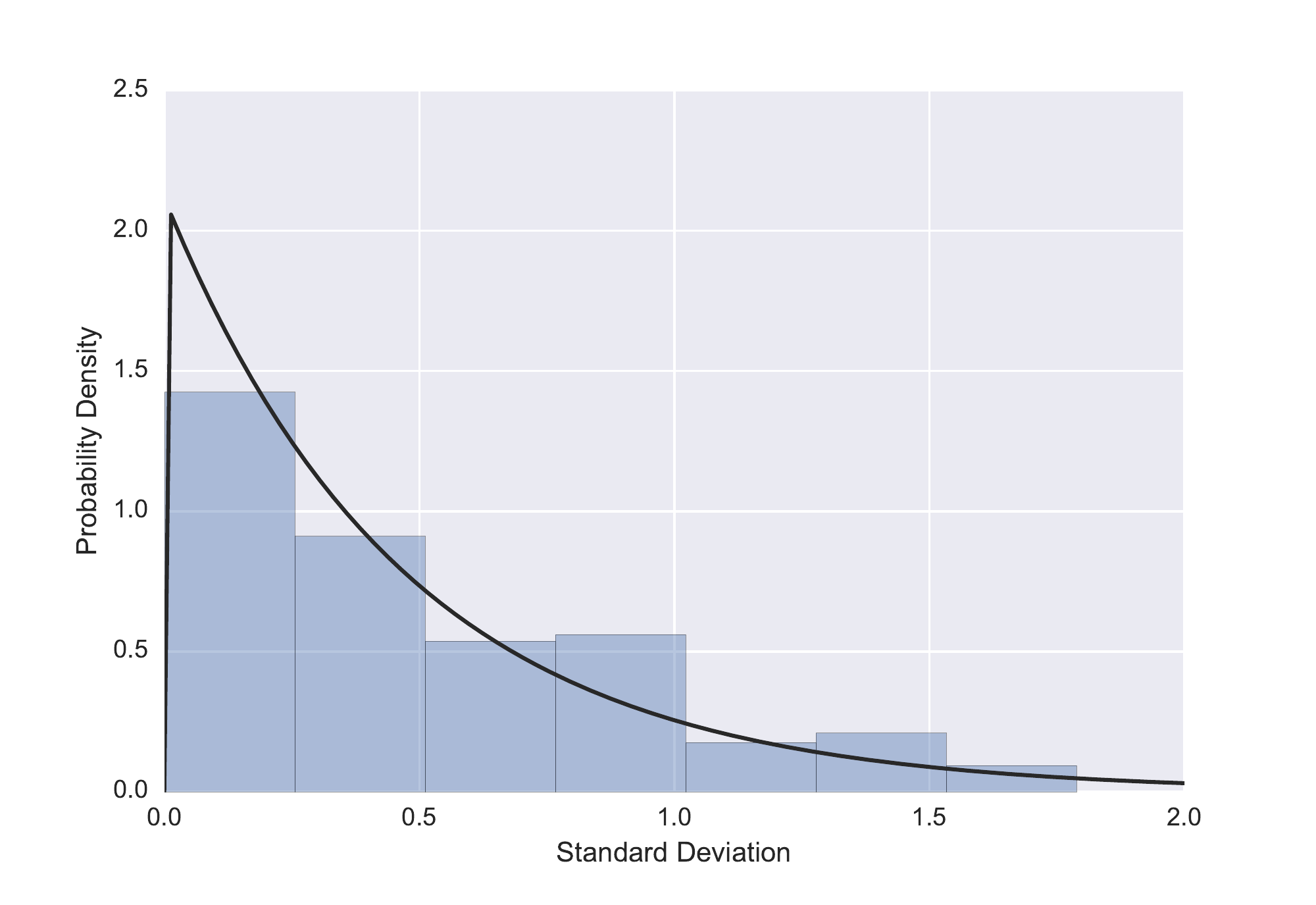}
        \caption{Distribution of variances emerged from repeated ratings within our experiment}
        \label{fig:VarDist}
\end{figure}

\paragraph{Proving Human Uncertainty}
For each of the user-item-pairs $R_{u,i}$, we compute the Gaussian ML-Parameters and consider the variances $\mathbb{V}(R_{u,i})$ as representations of the Human Uncertainty. In our experiment, only few tensor slices contain constant ratings and hence lead to a vanishing variance. Performing an item-wise analysis, the fraction of tensor slices with non-zero variance ranges from 50 to 90\% that is, only every second participant is able to reproduce its own decisions for the best case. For the worst case, only one out of ten participants is able to precisely reproduce a rating.
Figure \ref{fig:VarDist} depicts the distribution of variances emerged from repeated ratings within our experiment.
We observe that the overall variance follows an exponential distribution $\mathbb{V}\sim\operatorname{Exp}(\lambda)$ with parameter $\lambda=2.11$. This power-law distribution literally means, that many users have a low degree of uncertainty while only a few users have a very high degree of uncertainty.

\subsection{The Magic Barrier}
Figure \ref{fig:SensA} shows that the expected value of the Magic Barrier depends solely on the Human Uncertainty.
For our five-star scale as well as its minimum and maximum variances, the expectation should - when equation \ref{eq:MBapp} holds - be located in the interval $[0.40 \, ;\, 1.55]$. In the case of our experiment, we have $N=213$ rating distributions with non-vanishing variance. It is clear from Figure \ref{fig:SensB} that for this sample size, the distribution of the Human Uncertainty has a large impact on the variance of the Magic Barrier. According to equation\ref{eq:MBapp}, the variance of the Magic Barriers should be found in the interval $[0.0008  \, ;\, 0.0113]$.

\begin{figure}[b]
        \includegraphics[width=.9\linewidth]{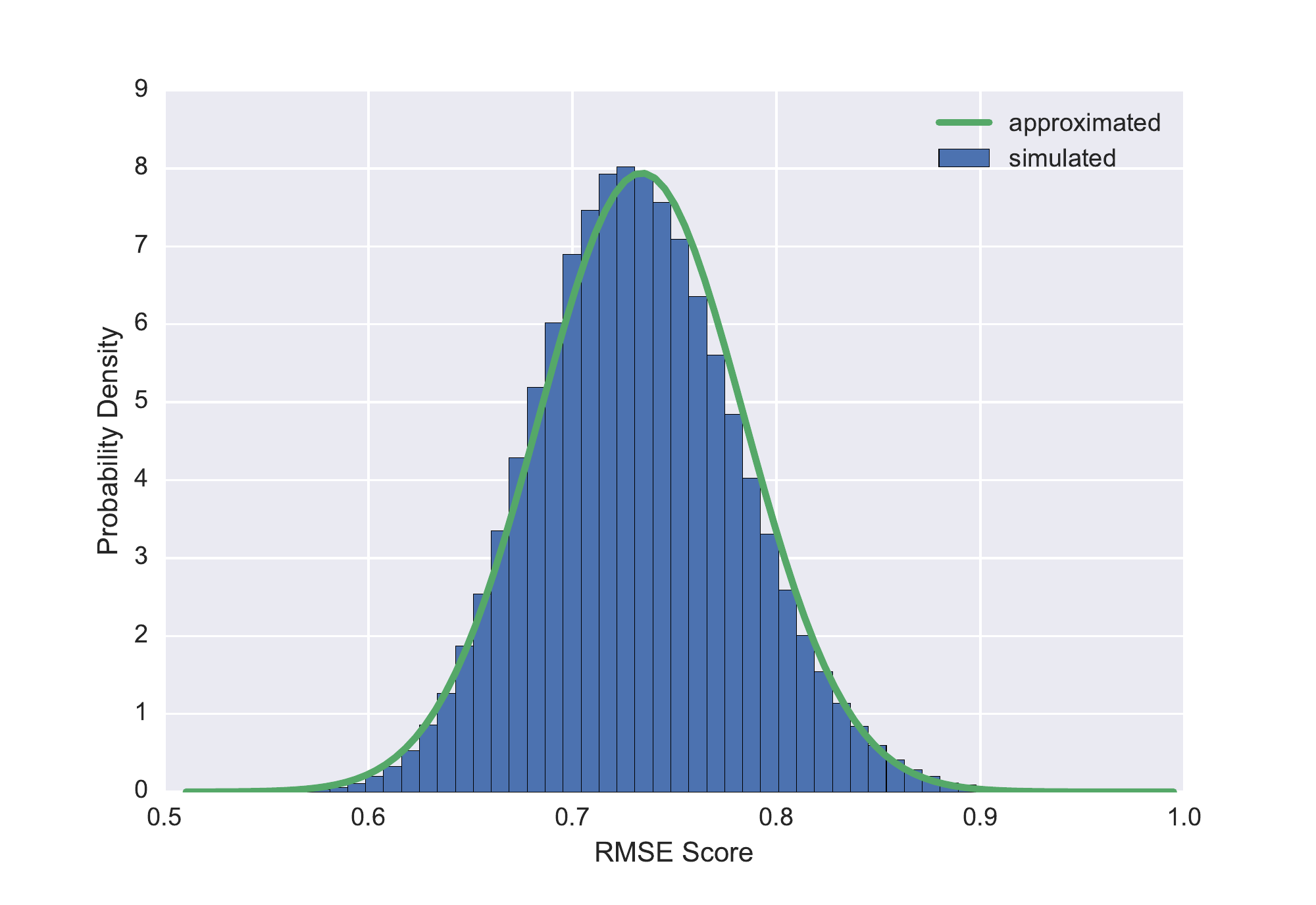}
        \caption{Visual comparison of simulated and approximation Magic Barrier based on experimental data records}
        \label{fig:ExpBarrierComp}
\end{figure}

On the basis of our data record, the simulation and approximation lead to well matching expectations (ca. $ 0.733$) and variances (ca. $0.003$) for the Magic Barrier. It is apparent, that the true values are located near the lower bound of the previously estimated intervals. This can be explained by the power-law distribution, i.e. a lot of variances are near the minimum and only a few have got higher extents. 
The difference of expectations is about $0.2\%$ while the difference of variances is about $1.2\%$. The matching between the simulated and the assumed data model of a Gaussian can be clearly confirmed in Figure \ref{fig:ExpBarrierComp}. The corresponding normed JSD is $0.05$. 

\subsection{Application}

\paragraph{Implicit Impact on Recommender Assessment}
So far we have only discussed the explicit impact on the assessment of recommender systems, that is: How likely is it that a system can still be improved, just before the RMSE solely depends on Human Uncertainty itself. Now we want to investigate the implicit influence, which affects any recommender comparison, even if the corresponding RMSE distributions are not directly overlapping with the Magic Barrier.

In doing so, we generate two copies of the Magic Barrier (as the optimal recommender). Each of these copies is gradually distorted by adding artificial noise to their predictors in such a way that the relative noise difference of both copies remain constant. By increasing the noise for both copies whilst keeping their relative difference constant, we generate an offset (distance from the Magic Barrier). This offset is plotted against the probabilities of error when using the traditional point-paradigm ranking, which is given by the generalisation of equation \ref{eq:BarrierInterference}. 
\begin{figure}[t]
        \includegraphics[width=\linewidth]{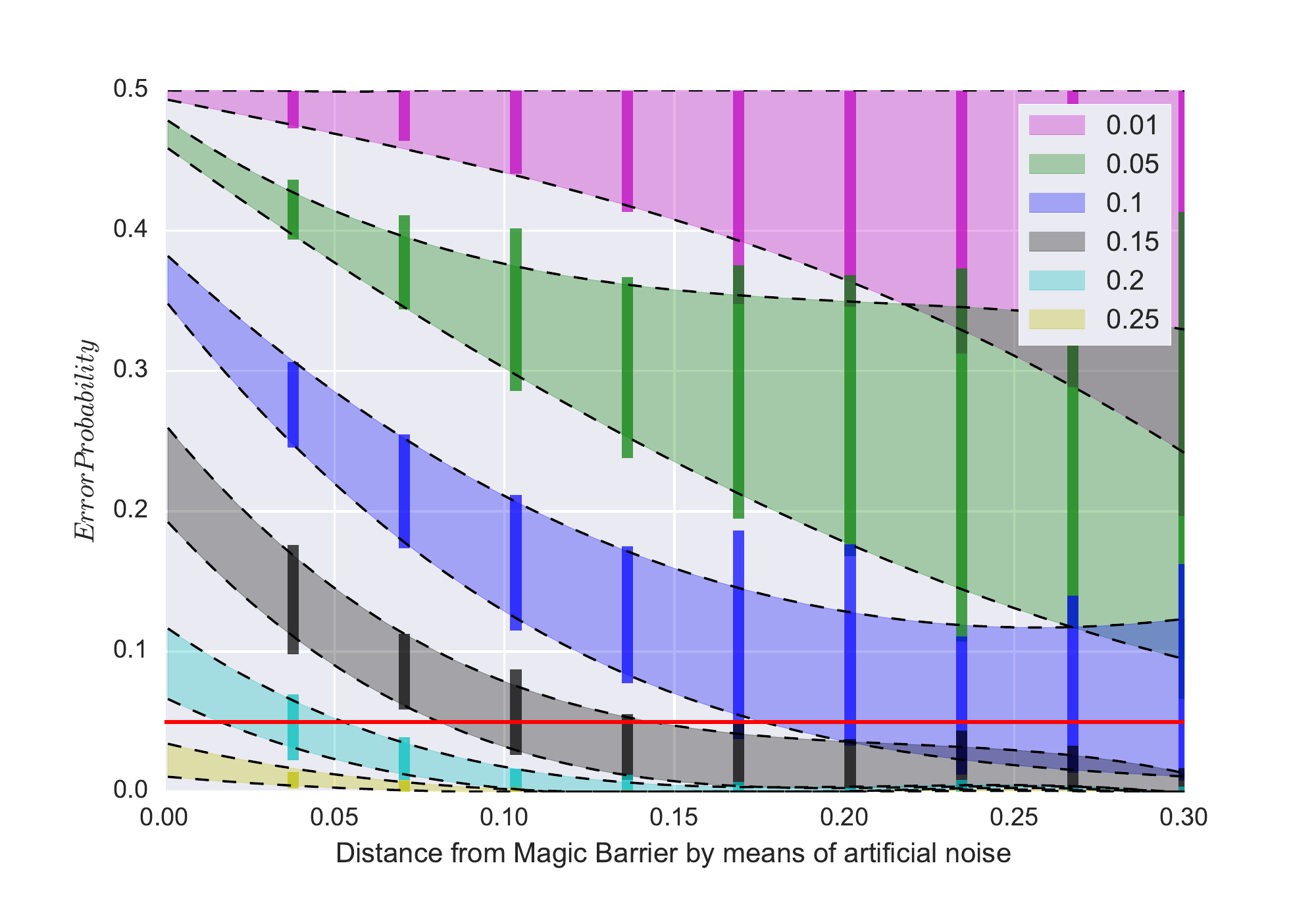}
        \caption{Error Probabilities for a point-paradigm ranking of systems with constant RMSE difference according to their overall distance to the Magic Barrier.}
        \label{fig:BarrierDistance}
\end{figure}
Figure \ref{fig:BarrierDistance} depicts the family of curves, mapping the distance from the Magic Barrier to the corresponding error probabilities. This distance (x-axis) represents the overall quality of a system, i.e. the larger this quantity, the worse the prediction quality. The colours encode the relative difference $\Delta$ of two recommender systems among each other. For the green curve (representing 10\% noise of difference), an $x$-value of $0.15$ means that system 1 has a noise of 15\% whereas system 2 has a noise of 25\%. The corresponding $y$-value indicates the error probability for ranking both systems using the traditional point-paradigm. It is apparent, that two recommender systems can not be brought into a ranking order without considerable error probability if their relative difference is less than 15\%, regardless of their basic prediction quality.

As a result, we recognise the following: The distance from the Magic Barrier has a great influence on the overlaps in two constantly different recommender systems, i.e. for a fixed difference in prediction quality, they can be distinguished much better if they are bad systems, rather than good ones. On the contrary, the better a system becomes, the more improvement does a revision need, in order to be detected with statistical evidence. This basically means that a recommender system within a repeated process of improvement will certainly reach a prediction quality so that there is probably no sufficiently large amount of optimisation left, in order to distinguish further improvements from the old system with statistical evidence. This convergence is actually the true nature of the Magic Barrier, which could not have been shown without switching perspectives to the distribution-paradigm.

\paragraph{Transferability: The Netflix Prize}
Unfortunately, existing records have not gathered Human Uncertainty.
Therefore, we examine the possibility of applying the findings of our experiment to such data records. To this end, we assume the distribution of Human Uncertainty, emerged from our experiment, to be valid for a larger number of ratings. Under this condition, we will examine possible consequences on Netflix Prize as an example.

The Netflix test record consists of $N = 2.8\cdot 10^6$ ratings in total. For each of these ratings, we randomly assign a variance drawn from the Pareto distribution in Figure \ref{fig:VarDist}. According to this data, the Magic Barrier can be estimated to $\mathcal{MB}\sim\mathcal{N}(0.6687, 0.0007)$. Even though the standard deviation seems small, it is still in the range of Netflix's rounding accuracy of four decimal places. To estimate whether the contest winner \cite{netflixleaderboard} might interfere with the Magic Barrier, we use the simplification of Equation \ref{eq:criterion}. Since $\mathbb{E}[\text{RMSE}_{best}] - \mathbb{E}[\mathcal{MB}] =  0.8567 -  0.6687 = 0.1880$ is greater than
$6\mathbb{V}[\text{MB}]^{1/2} = 0.1587$, it can be assumed that the Magic Barrier has not yet been reached. In fact, there is still the potential for about 20\% of improvement when taking the winner as reference.

\section{Discussion and Conclusions}

\paragraph*{Discussion}
In our experiment, the existence of Human Uncertainty is proven and it has been shown that it corresponds to a power-law distribution, i.e. there are many users having a small variance and there are only a few users having a large variance.
This implies the existence of an offset within every prediction quality metric that emerges from Human Uncertainty, the so-called Magic Barrier. Having several recommender systems whose RMSEs, for example, are lower than this Magic Barrier, every repetition of the rating proceeding would very likely result into rearrangements of the ranking order, i.e. a reliable ranking can not be built. 

In this article, we have lifted an existing theory of this Magic Barrier into a completely probabilistic methodology, providing a generalisation for any quality related metric. Our estimation provides processing of big data in little time while additionally being very precise. With our probabilistic approach, the true nature of the Magic Barrier can be demonstrated:
When approaching the Magic Barrier, the distinguishability of many recommender systems automatically decreases, supporting the idea of one equivalence class of optimal systems. 
Likewise, essential properties of the Magic Barrier have been revealed, for example, the expectation does not change for a higher number of ratings. In contrast, the variance even decreases for an additional number of uncertain ratings and allows to locate the Magic Barrier more precisely. Finally, we have demonstrated the possibility to transfer our results onto other data records in order to make careful predictions of possible interference.

\paragraph*{Conclusion}
What are the consequences for the assessment of recommender systems in general? The essence of our contribution is the revelation of the following problems:
\begin{enumerate}
\item People are not able to tell us what they really mean.
\item Human Uncertainty creates a barrier from which below any assessment results are just random.
\item This barrier also implicitly influences recommender assessments; the better our systems become, the more indistinguishable they become.
\end{enumerate}
At this point it must be said that these problems are not grounded in this new perspective presented here, but have always been present in data analysis. The approach used in this contribution is just able to make these problems visible. Furthermore, these problems do not only occur within our experiments, but have also been proven by other authors in different situations of user feedback. This may have far-reaching consequences, especially in the area of the recommender systems, when the selection of a supposedly better system is a monetary decision. For example, financial resources may be invested in improving a system but the improvements achieved are purely random, which remains unnoticed.

For this reason, it becomes crucial to further examine the extent of impact of Human Uncertainty within this field of research. It is also necessary to find proper solutions for these problems, e.g. designing sophisticated mechanisms to identify uncertainty and developing novel strategies to efficiently deal with it. This naturally involves research that connects the fields of behavioural decision making, cognitive psychology and recommender systems to create interdisciplinary synergy effects. We will continue to address these issues in further research.

\newpage

%\textbf{psychologische barrier}
%\begin{itemize}
%\item wir sind auch nciht voll probabilistisch
%\item eigentlich lernen recsys vom menschen das verhalten, also können sie maximal die verteilung erlernen
%\item ein prediktor kann also nie certain sein sondern hat dieselbe uncertainty wie der user.
%\item damit sind prediktoren aber auch zufallsvariablen
%\item der beste reccommender ist dann derjenice, der genau die rating verteilung vorraussagt
%\item Betrachten wir also (X-X)**2, so ergibt sich eine weitere (psychologische) magic barrier
%\item diese überlagert die gesamte top10 von netflix
%\item um also die systeme wirklcih zu verbessern, muss die erlernete uncertainty vom system reduziert werden
%\item wir müssen besser wissen was der mensch will als der mensch selber
%\item wir müssen seine unsicherheit dafür durchdringen und herausrechnen können
%\item zukünftige recommender müssen also psychologen sein
%\item Further Research: Auf in die Psychometrie!
%\end{itemize}

%\nocite{*}

%		8 pages + references

\vfill\eject
\bibliographystyle{acm}
\bibliography{Literature} 

\end{document}